\documentclass{PoS-pdf}
\pdfoutput=1
\usepackage{amstext}
\newcommand{\MS}{\ensuremath{\overline{\mbox{MS}}}}
\newcommand{\Tr}{\mathop{\mathrm{Tr}}\nolimits}

\title{Simultaneous decoupling of bottom and charm quarks}
\ShortTitle{Simultaneous decoupling of bottom and charm quarks}
\author{\speaker{Andrey Grozin}%
\thanks{Permanent address: Budker Institute of Nuclear Physics SB RAS, Novosibirsk.},
Maik H\"oschele, Jens Hoff, and Matthias Steinhauser\\
Institut f\"ur Theoretische Teilchenphysik,
Karlsruher Institut f\"ur Technologie, Karlsruhe\\
E-mail: \email{A.G.Grozin@inp.nsk.su},
\email{hoeschele@particle.uni-karlsruhe.de},
\email{jens@particle.uni-karlsruhe.de},
and \email{matthias.steinhauser@kit.edu}}
\abstract{Parameters and light fields of the QCD Lagrangian
with two heavy flavours, $b$ and $c$, are related
to those in the low-energy effective theory without these flavours,
to three-loop accuracy taking into account the exact dependence on $m_c/m_b$.
Similar relations for bilinear quark currents are also considered.}
\FullConference{Loops and Legs in Quantum Field Theory -- 11th DESY Workshop on Elementary Particle Physics\\
April 15--20, 2012\\
Wernigerode, Germany}

\begin{document}

\section{Introduction}
\label{S:Intro}

\begin{figure}[h]
\begin{center}
\begin{picture}(36,36)
\put(18,18){\makebox(0,0){\includegraphics{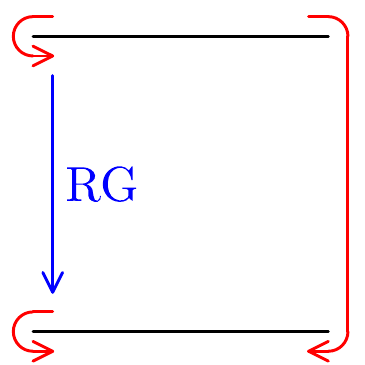}}}
\put(18,34){\makebox(0,0){$m_b$}}
\put(18,1){\makebox(0,0){$m_c$}}
\end{picture}
\end{center}
\caption{Two-step and one-step decoupling.}
\label{F:21}
\end{figure}

We consider QCD with $n_l$ light flavours,
$n_c$ flavours with mass $m_c$,
and $n_b$ flavours with mass $m_b$
($m_{c,b}\gg\Lambda_{\text{QCD}}$,
$m_c/m_b\lesssim1$;
the total number of flavours is $n_f=n_l+n_c+n_b$).
At low energies $\ll m_{c,b}$ it is appropriate to use
the effective theory without both $b$ and $c$.
It has the ordinary QCD Lagrangian (with re-defined fields and parameters)
plus $1/m_{c,b}^n$ corrections (higher-dimensional operators).
Operators in full QCD (e.g., bilinear quark currents)
are also expressed as series in $1/m_{c,b}$ via operators in the effective theory.
Traditionally~\cite{BW:82,LRV:95,CKS:98,SS:06,CKS:06}, one first decouples $b$ quarks,
producing an intermediate effective theory;
then one decouples $c$ quarks (Fig.~\ref{F:21}).
Power corrections $(m_c/m_b)^n$ are neglected in this approach.
Its advantage is the possibility to sum leading, next-to-leading, etc.,
powers of $\log(m_b/m_c)$;
however, this logarithm is not really large.

Alternatively~\cite{GHHS:11}, one can decouple both $b$ and $c$ in a single step
(Fig.~\ref{F:21}).
This can be done at some fixed order in $\alpha_s$,
renormalization group summation of powers of $\log(m_b/m_c)$
is not possible.
However, now coefficients of perturbative series can be calculated
as exact functions of $m_c/m_b$.
A non-trivial dependence on $m_c/m_b$ starts from three loops,
and the first power correction $(m_c/m_b)^2 (\alpha_s/\pi)^3$
may be of the same order as recently computed~\cite{SS:06,CKS:06}
four-loop corrections $(\alpha_s/\pi)^4$.
It is possible to combine advantages of both approaches
by RG summing powers of $\log(m_b/m_c)$ in the leading power term~\cite{GHHS:11}.

\section{Decoupling for fields and parameters of the Lagrangian}
\label{S:L}

The bare fields in the two theories are related by
\begin{equation}
A_0^{(n_l)} = \left(\zeta_A^0\right)^{1/2} A_0^{(n_f)}\,,\quad
c_0^{(n_l)} = \left(\zeta_c^0\right)^{1/2} c_0^{(n_f)}\,,\quad
q_0^{(n_l)} = \left(\zeta_q^0\right)^{1/2} q_0^{(n_f)}\,,
\label{L:f0}
\end{equation}
where power corrections are neglected.
Similarly, the bare parameters of the Lagrangians are related by
\begin{equation}
\alpha_{s0}^{(n_l)} = \zeta_\alpha^0 \alpha_{s0}^{(n_f)}\,,\quad
a_0^{(n_l)} = \zeta_A^0 a_0^{(n_f)}\,,\quad
m_{q0}^{(n_l)} = \zeta_m^0 m_{q0}^{(n_f)}\,.
\label{L:p0}
\end{equation}
The \MS{} renormalized fields and parameters are related by
\begin{equation}
\alpha_s^{(n_l)}(\mu') = \zeta_\alpha(\mu',\mu) \alpha_s^{(n_f)}(\mu)\,,\quad
A^{(n_l)}(\mu') = \zeta_A^{1/2}(\mu',\mu) A^{(n_f)}(\mu)\,,
\label{L:fp}
\end{equation}
where
\begin{eqnarray}
&&\zeta_\alpha(\mu',\mu) =
\frac{Z_\alpha^{(n_f)}\left(\alpha_s^{(n_f)}(\mu)\right)}%
{Z_\alpha^{(n_l)}\left(\alpha_s^{(n_l)}(\mu')\right)}
\zeta_\alpha^0\left(\alpha_{s0}^{(n_f)}\right)\,,
\nonumber\\
&&\zeta_A(\mu',\mu) =
\frac{Z_A^{(n_f)}\left(\alpha_s^{(n_f)}(\mu),a^{(n_f)}(\mu)\right)}%
{Z_A^{(n_l)}\left(\alpha_s^{(n_l)}(\mu'),a^{(n_l)}(\mu')\right)}
\zeta_A^0\left(\alpha_{s0}^{(n_f)},a_0^{(n_f)}\right)\,,
\label{L:ren}
\end{eqnarray}
and similarly for other decoupling coefficients.

Let's consider, for example, the gluon field.
In the full QCD ($n_f$ flavours), the bare field is related
to the field in the on-shell scheme by
$\displaystyle A_0^{(n_f)} = Z_A^{\text{os}(n_f)} A_{\text{os}}^{(n_f)}$,
where $Z_A^{\text{os}(n_f)}=1/(1+\Pi_A^{(n_f)}(0))$,
and $\Pi_A^{(n_f)}(0)$ contains at least one heavy-quark loop.
Similarly, in the effective theory ($n_l$ flavours)
$\displaystyle A_0^{(n_l)} = Z_A^{\text{os}(n_l)} A_{\text{os}}^{(n_l)}$,
where $Z_A^{\text{os}(n_l)}=1/(1+\Pi_A^{(n_l)}(0))$,
and $\Pi_A^{(n_l)}(0)=0$ because it has no scale.
Gluon propagators renormalized in the on-shell scheme in both theories
are equal to the free one near the mass shell,
and hence $A_{\text{os}}^{(n_f)}=A_{\text{os}}^{(n_l)}$ up to power corrections.
Therefore
\begin{equation}
\zeta_A^0 = \frac{Z_A^{\text{os}(n_l)}}{Z_A^{\text{os}(n_f)}}
= 1 + \Pi_A^{(n_f)}(0)\,,
\quad\text{and}\quad
\zeta_A(\mu',\mu) = \frac{Z_A^{(n_f)}}{Z_A^{(n_l)}}
\frac{Z_A^{\text{os}(n_l)}}{Z_A^{\text{os}(n_f)}}\,.
\label{L:A0}
\end{equation}
Other fields are treated in the same way.
The bare decoupling coefficient for light-quark masses is given by
\begin{equation}
\zeta_m^0 = Z_q^{\text{os}} \left[1 - \Sigma_S(0)\right]\,,
\label{L:m}
\end{equation}
where the light-quark self-energy is
$\Sigma(p)=\rlap{\hspace{0.8mm}/}p \Sigma_V(p^2)+m_{q0} \Sigma_S(p^2)$.

In order to find the decoupling for $\alpha_s$,
one has to consider some vertex function:
$A\bar{c}c$, $A\bar{q}q$, or $AAA$.
They are expanded in their external momenta,
and only the leading non-vanishing terms are retained.
In the low-energy theory they get no loop corrections,
and are given by the tree-level vertices of dimension-4 operators in the Lagrangian.
In full QCD they are equal to the  tree-level vertices times $1+\Gamma_i$,
where loop corrections $\Gamma_i$ contain at least one heavy-quark loop%
\footnote{For $A\bar{q}q$ at the zeroth order in external momenta,
there is the only structure $\gamma^\mu$.
For $A\bar{c}c$ at the first order we shall prove this a little later.
The $AAA$ vertex at the first order in its external momenta can have,
in addition to the tree-level structure, one more structure:
$d^{a_1 a_2 a_3}(g^{\mu_1 \mu_2}k_3^{\mu_3}+\mbox{cycle})$;
however, the Slavnov--Taylor identity
$\langle T\{\partial^\mu A_\mu(x),\partial^\nu A_\nu(y),\partial^\lambda A_\lambda(z)\}\rangle=0$
leads to $\Gamma^{a_1 a_2 a_3}_{\mu_1 \mu_2 \mu_3} k_1^{\mu_1} k_2^{\mu_2} k_3^{\mu_3} = 0$,
thus excluding this second structure.}.
Then
\begin{equation}
\zeta_\alpha^0(\alpha_{s0}^{(n_f)})
= \left(1+\Gamma_{A\bar{c}c}\right)^2 \left(Z_c^{\text{os}}\right)^2 Z_A^{\text{os}}
= \left(1+\Gamma_{A\bar{q}q}\right)^2 \left(Z_q^{\text{os}}\right)^2 Z_A^{\text{os}}
= \left(1+\Gamma_{AAA}\right)^2 \left(Z_A^{\text{os}}\right)^3\,.
\label{L:alpha}
\end{equation}

\begin{figure}[ht]
\begin{center}
\includegraphics{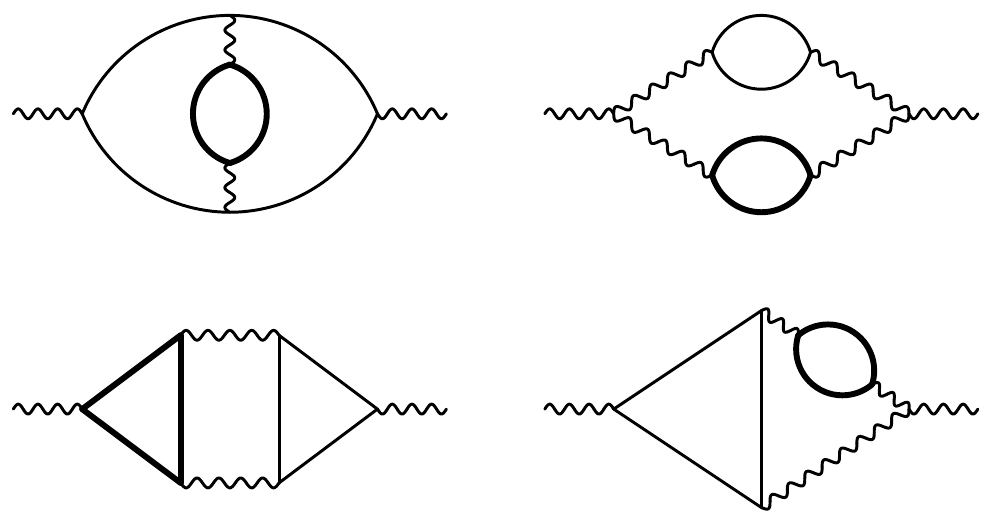}
\end{center}
\caption{Some contributions to the gluon self-energy
with both $b$ (thick) and $c$ (thin) loops.}
\label{F:Glue}
\end{figure}

The gluon self-energy up to three loops has the structure
\begin{eqnarray}
&&\Pi_A(0) = \frac{1}{3}
\left( n_b m_{b0}^{-2\varepsilon} + n_c m_{c0}^{-2\varepsilon} \right)
T_F \frac{\alpha_{s0}^{(n_f)}}{\pi} \Gamma(\varepsilon)
+ P_h \left( n_b m_{b0}^{-4\varepsilon} + n_c m_{c0}^{-4\varepsilon} \right)
T_F \left(\frac{\alpha_{s0}^{(n_f)}}{\pi} \Gamma(\varepsilon)\right)^2
\nonumber\\
&&\hphantom{\Pi_A(0)={}} +
\biggl[ \left(P_{hg} + P_{hl} T_F n_l\right)
\left( n_b m_{b0}^{-6\varepsilon} + n_c m_{c0}^{-6\varepsilon} \right)
+ P_{hh} T_F \left( n_b^2 m_{b0}^{-6\varepsilon} + n_c^2 m_{c0}^{-6\varepsilon} \right)
\nonumber\\
&&\hphantom{\Pi_A(0)={}+\biggl[\biggr.} +
P_{bc}\left(\frac{m_{c0}}{m_{b0}}\right)
T_F n_b n_c \left(m_{b0} m_{c0}\right)^{-3\varepsilon}
\biggr] T_F \left(\frac{\alpha_{s0}^{(n_f)}}{\pi} \Gamma(\varepsilon)\right)^3 + \cdots\,.
\label{L:PiA}
\end{eqnarray}
The contribution involving both $b$ and $c$ first appears at three loops
(Fig.~\ref{F:Glue}) and satisfies
\begin{equation}
P_{bc}(x^{-1}) = P_{bc}(x)\,,\quad
P_{bc}(1) = 2 P_{hh}\,,\quad
P_{bc}^{\text{hard}}(x\to0) \to P_{hl} x^{3\varepsilon}\,,
\end{equation}
where $P_{bc}^{\text{hard}}$ is the contribution of the hard region
(all loop momenta $\sim m_b$).

\begin{figure}[ht]
\begin{center}
\includegraphics{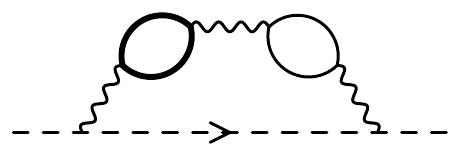}
\end{center}
\caption{The contribution to the ghost self-energy
with both $b$ and $c$ loops.}
\label{F:Ghost}
\end{figure}

The ghost self-energy up to three loops has the structure
\begin{eqnarray}
&&\Pi_c(0) =
C_h \left(n_b m_{b0}^{-4\varepsilon} + n_c m_{c0}^{-4\varepsilon}\right)
C_A T_F \left(\frac{\alpha_{s0}^{(n_f)}}{\pi} \Gamma(\varepsilon)\right)^2
\nonumber\\
&&\hphantom{\Pi_c(0)={}} +
\biggl[ \left(C_{hg} + C_{hl}T_F n_l\right)
\left( n_b m_{b0}^{-6\varepsilon} + n_c m_{c0}^{-6\varepsilon} \right)
+ C_{hh}T_F \left( n_b^2 m_{b0}^{-6\varepsilon} + n_c^2 m_{c0}^{-6\varepsilon} \right)
\nonumber\\
&&\hphantom{\Pi_c(0)={}{}+\biggl[\biggr.} +
C_{bc}\left(\frac{m_{c0}}{m_{b0}}\right)
T_F n_b n_c \left(m_{b0} m_{c0}\right)^{-3\varepsilon}
\biggr] C_A T_F \left(\frac{\alpha_{s0}^{(n_f)}}{\pi} \Gamma(\varepsilon)\right)^3 + \cdots\,.
\label{L:Ghost}
\end{eqnarray}
The contribution with both $b$ and $c$ (Fig.~\ref{F:Ghost}) has similar properties.

The light-quark self-energy up to three loops has the structure
\begin{eqnarray}
&&\Sigma_V(0) =
V_h \left(n_b m_{b0}^{-4\varepsilon} + n_c m_{c0}^{-4\varepsilon}\right)
C_F T_F \left(\frac{\alpha_{s0}^{(n_f)}}{\pi} \Gamma(\varepsilon)\right)^2
\nonumber\\
&&\hphantom{\Sigma_V(0)={}} + \biggl[ \left(V_{hg} + V_{hl} T_F n_l\right)
\left( n_b m_{b0}^{-6\varepsilon} + n_c m_{c0}^{-6\varepsilon} \right)
+ V_{hh} T_F \left( n_b^2 m_{b0}^{-6\varepsilon} + n_c^2 m_{c0}^{-6\varepsilon} \right)
\nonumber\\
&&\hphantom{\Sigma_V(0)={}+\biggl[\biggr.} +
V_{bc}\left(\frac{m_{c0}}{m_{b0}}\right)
T_F n_b n_c \left(m_{b0} m_{c0}\right)^{-3\varepsilon}
\biggr] C_F T_F \left(\frac{\alpha_{s0}^{(n_f)}}{\pi} \Gamma(\varepsilon)\right)^3 + \cdots\,,
\end{eqnarray}
where $V_{bc}$ comes from the diagram similar to Fig.~\ref{F:Ghost};
$\Sigma_S(0)$ has the same structure.

We choose to use the ghost--gluon vertex for finding $\zeta_\alpha$.
When expanded in its external momenta up to linear terms,
it has the structure
\[
\raisebox{-7mm}{\begin{picture}(27,18)
\put(13.5,9.5){\makebox(0,0){\includegraphics{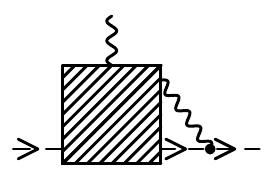}}}
\put(23.5,0){\makebox(0,0)[b]{$p$}}
\put(12.5,18){\makebox(0,0)[tl]{$\mu$}}
\put(22,4.5){\makebox(0,0)[bl]{$\nu$}}
\end{picture}}
= A^{\mu\nu} p_\nu\,,\quad
A^{\mu\nu}(0) = A g^{\mu\nu}\,,
\]
i.e., it is proportional to the tree vertex.
The leftmost vertex on the ghost line singles out
the longitudinal part of the gluon propagator:
\[
\raisebox{-7mm}{\begin{picture}(27,18)
\put(13.5,9.5){\makebox(0,0){\includegraphics{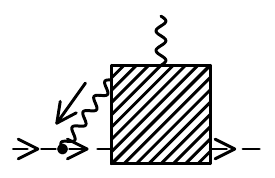}}}
\put(3.5,0){\makebox(0,0)[b]{$0$}}
\put(8.5,0){\makebox(0,0)[b]{$k$}}
\put(5,10){\makebox(0,0){$k$}}
\end{picture}}\quad
\sim k^\lambda\,.
\]
Therefore all corrections vanish in Landau gauge.
Those with a quark loop in the leftmost gluon line
vanish in any covariant gauge:
\[
\raisebox{-9mm}{\includegraphics{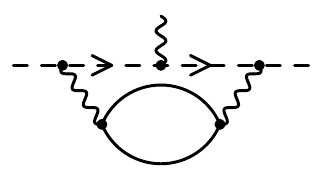}} =
\raisebox{-7mm}{\includegraphics{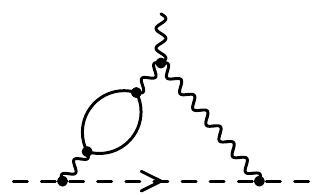}} = 0\,.
\]
Diagrams with quark triangles produce differences:
\begin{eqnarray*}
&&\raisebox{-7mm}{\includegraphics{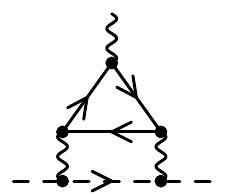}}
= a_0 \left[ \raisebox{-7mm}{\includegraphics{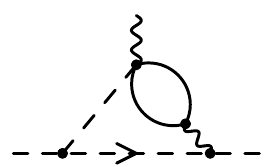}}
- \raisebox{-7mm}{\includegraphics{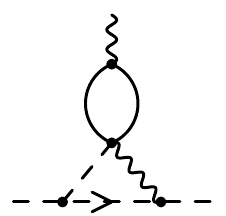}} \right]\,,\\
&&\raisebox{-7mm}{\includegraphics{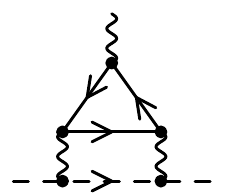}}
= a_0 \left[ \raisebox{-7mm}{\includegraphics{ghb.pdf}}
- \raisebox{-7mm}{\includegraphics{gha.pdf}} \right]\,,
\end{eqnarray*}
where those with a massless triangle vanish;
the remaining two give $[t^a,t^b]$.
Finally, the diagram with the three-gluon vertex
\[
\raisebox{-7mm}{\includegraphics{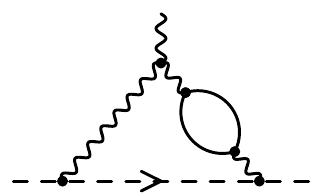}}
= a_0 \raisebox{-7mm}{\includegraphics{gha.pdf}}
\]
gives $[t^b,t^a]$ and cancels the previous one.
In result, the two-loop vertex exactly vanishes.
The three-loop vertex vanishes in Landau gauge;
it contains no diagrams with both $b$ and $c$ loops.

\begin{figure}[ht]
\begin{center}
\includegraphics{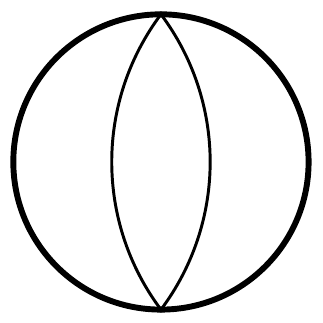}
\end{center}
\caption{The master integral.}
\label{F:Master}
\end{figure}

We used \texttt{FIRE}~\cite{S:08}
for reduction of two-scale three-loop Feynman integrals.
There are two master integrals: Fig.~\ref{F:Master}
and the same integral with a numerator.
They were considered in~\cite{BGSS:09,BKKWY:09}.
At $m_b=m_c$ they are not independent;
$\varepsilon$ expansion of the only master integral
has been considered in~\cite{B:92}.

The decoupling relation $\alpha_{s0}^{(n_l)} = \zeta_\alpha^0 \alpha_{s0}^{(n_f)}$
contains
\[
\zeta_\alpha^0 = 1
- \frac{1}{3} \left(n_b m_{b0}^{-2\varepsilon} + n_c m_{c0}^{-2\varepsilon}\right)
T_F \frac{\alpha_{s0}^{(n_f)}}{\pi} \Gamma(\varepsilon)
+ \cdots\,.
\]
We re-express the right-hand side via renormalized quantities
\begin{eqnarray*}
&&\frac{\alpha_{s0}^{(n_f)}}{\pi} \Gamma(\varepsilon) =
\frac{\alpha_s^{(n_f)}(\mu)}{\pi \varepsilon}
Z_\alpha^{(n_f)}\left(\alpha_s^{(n_f)}(\mu)\right)
e^{\gamma_E \varepsilon} \Gamma(1+\varepsilon)
\mu^{2\varepsilon}\,,\\
&&m_{b0} = Z_m^{(n_f)}\left(\alpha_s^{(n_f)}(\mu)\right) m_b(\mu)\,,\quad
m_{c0} = Z_m^{(n_f)}\left(\alpha_s^{(n_f)}(\mu)\right) m_c(\mu)\,.
\end{eqnarray*}
It is convenient to use $\mu=\bar{m}_b$ which is defined
as the root of the equation $m_b(\bar{m}_b) = \bar{m}_b$.
We obtain $\alpha_{s0}^{(n_l)}$ expressed via $\alpha_s^{(n_f)}(\bar{m}_b)$,
$\bar{m}_b$, and $m_c(\bar{m}_b)$.
Inverting the series
\[
\frac{\alpha_{s0}^{(n_l)}}{\pi} \Gamma(\varepsilon) =
\frac{\alpha_s^{(n_l)}(\mu')}{\pi \varepsilon}
Z_\alpha^{(n_l)}\left(\alpha_s^{(n_l)}(\mu')\right)
e^{\gamma_E \varepsilon} \Gamma(1+\varepsilon)
\mu^{\prime2\varepsilon}\,,
\]
we obtain $\alpha_s^{(n_l)}(\mu')$.
It is convenient to use $\mu'=m_c(\bar{m}_b)$,
then the result is expressed via $\alpha_s^{(n_f)}(\bar{m}_b)$ and
\[
x = \frac{m_c(\bar{m}_b)}{\bar{m}_b}\,.
\]
The procedure for gauge-dependent decoupling coefficients
(e.g., $\zeta_A$) is similar but a little more lengthy.

\section{Decoupling for bilinear quark currents}
\label{S:j}

Until now we considered the fields and parameters of the Lagrangian.
Other operators in full QCD also can be expressed
via operators in the effective theory.
Here we shall discuss bilinear quark currents~\cite{G:12}.

\subsection{Light non-singlet currents}

First we consider light-quark currents
\begin{equation}
j_{n0}^{(n_f)} = \bar{q}_0^{(n_f)} \Gamma_n \tau q_0^{(n_f)}
= Z_n^{(n_f)}(\alpha_s^{(n_f)}(\mu)) j_n^{(n_f)}(\mu)\,,
\label{j:lsn}
\end{equation}
where $\Gamma_n = \gamma^{[\mu_1} \cdots \gamma^{\mu_n]}$
is the antisymmetrized product of $n$ $\gamma$-matrices,
and $\tau$ is a non-singlet flavour matrix.
Their anomalous dimensions are known up to three loops~\cite{G:00},
and $n$ appears in the formula as a symbolic parameter.

These currents can be expressed via operators in the $n_l$ flavour theory:
\begin{equation}
j_n^{(n_f)}(\mu) = \zeta_n^{-1}(\mu',\mu) j_n^{(n_l)}(\mu')
+ \mathcal{O}(1/m_{c,b}^2)\,.
\label{j:dec}
\end{equation}
It is convenient first to find the bare decoupling
$j_{n0}^{(n_f)} = \bigl(\zeta_n^0\bigr)^{-1} j_{n0}^{(n_l)}$
(neglecting power corrections), then
\begin{equation}
\zeta_n(\mu',\mu) =
\frac{Z_n^{(n_f)}(\alpha_s^{(n_f)}(\mu))}{Z_n^{(n_l)}(\alpha_s^{(n_l)}(\mu'))}
\zeta_n^0\,.
\label{j:ren}
\end{equation}
The bare decoupling coefficient is obtained by matching on-shell matrix elements:
\begin{equation}
Z_q^{\text{os}(n_f)} \left( 1 + \Gamma_n^{(n_f)} \right)
= \bigl(\zeta_n^0\bigr)^{-1}
Z_q^{\text{os}(n_l)} \left( 1 + \Gamma_n^{(n_l)} \right)\,,
\label{j:match0}
\end{equation}
where $Z_q^{\text{os}(n_l)}=1$ and $\Gamma_n^{(n_l)}=0$ at zero external momenta.
Therefore,
\begin{equation}
\bigl(\zeta_n^0\bigr)^{-1} =
Z_q^{\mbox{\scriptsize os}}(\alpha_{s0}^{(n_f)},a_0^{(n_f)})
\left( 1 + \Gamma_n(\alpha_{s0}^{(n_f)},a_0^{(n_f)}) \right)\,.
\label{j:match}
\end{equation}
Decoupling of a single heavy flavour for light-quark non-singlet currents
up to three loops has been considered in~\cite{GSS:06}
($n$ appears in the result as a symbolic parameter).

\begin{figure}[ht]
\begin{center}
\begin{picture}(82,30)
\put(23,10){\makebox(0,0){\includegraphics{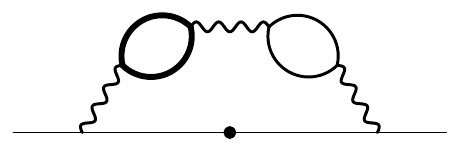}}}
\put(23,0){\makebox(0,0)[b]{a}}
\put(69,16.5){\makebox(0,0){\includegraphics{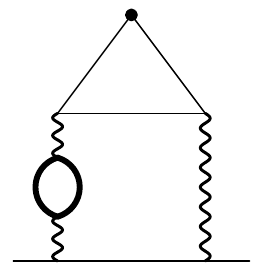}}}
\put(69,0){\makebox(0,0)[b]{b}}
\end{picture}
\end{center}
\caption{Vertices of light-quark currents
(the very thin line means light quark).}
\label{F:j}
\end{figure}

The vertex up to three loops has the structure
\begin{eqnarray}
&&\Gamma_n =
\Gamma_h \left(n_b m_{b0}^{-4\varepsilon} + n_c m_{c0}^{-4\varepsilon}\right)
C_F T_F \left(\frac{\alpha_{s0}^{(n_f)}}{\pi} \Gamma(\varepsilon)\right)^2
\nonumber\\
&&\hphantom{\Gamma_n={}}
+ \biggl[ \left(\Gamma_{hg} + \Gamma_{hl} T_F n_l\right)
\left( n_b m_{b0}^{-6\varepsilon} + n_c m_{c0}^{-6\varepsilon} \right)
+ \Gamma_{hh} T_F
\left( n_b^2 m_{b0}^{-6\varepsilon} + n_c^2 m_{c0}^{-6\varepsilon} \right)
\nonumber\\
&&\hphantom{\Gamma_n={}+\biggl[\biggr.} +
\Gamma_{bc}\left(\frac{m_{c0}}{m_{b0}}\right)
T_F n_b n_c \left(m_{b0} m_{c0}\right)^{-3\varepsilon}
\biggr] C_F T_F \left(\frac{\alpha_{s0}^{(n_f)}}{\pi} \Gamma(\varepsilon)\right)^3 + \cdots\,,
\label{j:Gamma}
\end{eqnarray}
where $\Gamma_{bc}$ comes from the diagram Fig.~\ref{F:j}a
($n$ appears in the result as a symbolic parameter), and
\begin{equation}
\Gamma_{bc}(x^{-1}) = \Gamma_{bc}(x)\,,\quad
\Gamma_{bc}(1) = 2 \Gamma_{hh}\,,\quad
\Gamma_{bc}^{\text{hard}}(x\to0) \to \Gamma_{hl} x^{3\varepsilon}\,.
\label{j:Gamma1}
\end{equation}

For the vector current ($n=1$)
\begin{equation}
\zeta_1^0 = 1\,,\quad
Z_1^{(n_f)} = Z_1^{(n_l)} = 1\,,\quad
\zeta_1(\mu',\mu) = 1
\label{j:1}
\end{equation}
to all orders%
\footnote{For example, if $\tau$ is diagonal with $\Tr\tau=0$,
then $\int j_1^\mu dS_\mu$ over all space
($dS_\mu$ is the area element of the hypersurface)
is a linear combination of differences of the numbers of quarks and antiquarks
of various flavours; these differences are integer numbers,
and do not change when we switch from the full QCD
to the effective theory.}.
For the scalar current ($n=0$)
\begin{equation}
\zeta_0^0 = \bigl(\zeta_m^0\bigr)^{-1}\,,\quad
\zeta_0(\mu',\mu) = \zeta_m^{-1}(\mu',\mu)\,.
\label{j:0}
\end{equation}

The current with $n=4$ is related to $n=0$,
and with $n=3$ --- to $n=1$.
We can define the renormalized currents with 't~Hooft--Veltman $\gamma_5$ as
\begin{equation}
\left(\bar{q} \gamma_5^{\text{HV}} \tau q\right)_\mu =
\frac{i}{4!} \varepsilon_{\alpha\beta\gamma\delta} j_4^{\alpha\beta\gamma\delta}(\mu)\,,\quad
\left(\bar{q} \gamma_5^{\text{HV}} \gamma_\delta \tau q\right)_\mu =
\frac{i}{3!} \varepsilon_{\alpha\beta\gamma\delta} j_3^{\alpha\beta\gamma}(\mu)\,,
\end{equation}
because the renormalized currents $j_4(\mu)$, $j_3(\mu)$ live in 4 dimensions.
The currents with ``anticommuting $\gamma_5$'' are~\cite{LV:91,L:93}
\begin{equation}
\left(\bar{q} \gamma_5^{\text{AC}} \tau q\right)_\mu =
Z_P(\alpha_s(\mu)) \left(\bar{q} \gamma_5^{\text{HV}} \tau q\right)_\mu\,,\quad
\left(\bar{q} \gamma_5^{\text{AC}} \gamma^\alpha \tau q\right)_\mu =
Z_P(\alpha_s(\mu)) \left(\bar{q} \gamma_5^{\text{HV}} \gamma^\alpha \tau q\right)_\mu\,,
\label{j:Larin}
\end{equation}
where the finite renormalization factors $Z_{P,A}$ are tuned
to make the anomalous dimensions of these currents equal to those
of $j_0$, $j_1$ (the last anomalous dimension is 0).
The anticommuting $\gamma_5$ does not influence decoupling coefficients,
and therefore
\begin{equation}
\frac{\zeta_4(\mu',\mu)}{\zeta_0(\mu',\mu)} =
\frac{Z_P^{(n_f)}(\alpha_s^{(n_f)}(\mu))}{Z_P^{(n_l)}(\alpha_s^{(n_l)}(\mu'))}\,,\quad
\frac{\zeta_3(\mu',\mu)}{\zeta_1(\mu',\mu)} =
\frac{Z_A^{(n_f)}(\alpha_s^{(n_f)}(\mu))}{Z_A^{(n_l)}(\alpha_s^{(n_l)}(\mu'))}\,.
\label{j:43}
\end{equation}
These ratios do not depend on $m_c/m_b$.

\subsection{Light singlet currents}

Now we shall consider flavour-singlet light-quark currents ($\tau=1$).
In addition to diagrams discussed above,
their vertex functions contain diagrams where the light quark
emitted in the current vertex returns to the same vertex.
These diagrams vanish for even $n$;
therefore, we have to consider $n=1$ and 3.
The decoupling coefficient for the vector current is exactly 1.
The anomalous dimension of the flavour-singlet $j_3$
has been calculated up to three loops in~\cite{L:93,CK:93}.
As compared to the non-singlet case, there is an additional vertex diagram
Fig.~\ref{F:j}b~\cite{CT:93,LRV:95}; it has the structure
\begin{equation}
A_s C_F T_F^2 n_l \left(n_b m_{b0}^{-6\varepsilon} + n_c m_{c0}^{-6\varepsilon}\right)
\left(\frac{\alpha_{s0}^{(n_f)}}{\pi} \Gamma(\varepsilon)\right)^3\,.
\end{equation}

\subsection{Heavy currents}

\begin{figure}[ht]
\begin{center}
\begin{picture}(62,30)
\put(13,14){\makebox(0,0){\includegraphics{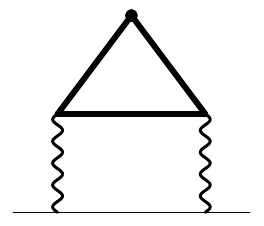}}}
\put(13,0){\makebox(0,0)[b]{a}}
\put(49,16.5){\makebox(0,0){\includegraphics{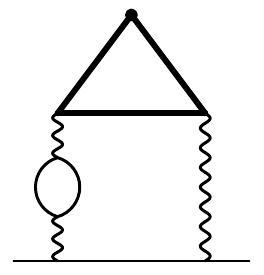}}}
\put(49,0){\makebox(0,0)[b]{b}}
\end{picture}
\end{center}
\caption{Vertices of $b$-quark currents.}
\label{F:h}
\end{figure}

Finally, we consider $b$-quark currents
(results for $c$-quark currents can be obtained by the obvious substitution).
They can produce flavour-singlet light-quark currents.
The number of $\gamma$ matrices on the light-quark line is always odd,
hence we should consider the currents with $n=1$ and 3.
The vector current simply becomes 0 in the effective theory,
up to power corrections.
The current with $n=3$ becomes the flavour-singlet light-quark current:
\begin{equation}
j_{b0}=\bar{b}_0\gamma^{[\alpha}\gamma^{\beta}\gamma^{\gamma]}b_0\,,\quad
j_{q0}=\bar{q}_0\gamma^{[\alpha}\gamma^{\beta}\gamma^{\gamma]}q_0\,,\quad
j_{b0}^{n_f} = \zeta_A^0 j_{q0}^{n_l}\,,
\label{j:b}
\end{equation}
where $\zeta_A^0$ starts from two loops (Fig.~\ref{F:h}a).
Decoupling of the single heavy flavour ($b$) has been calculated
up to three loops in~\cite{CK:93,CT:93}.
When decoupling both $b$ and $c$,
\begin{eqnarray}
&&\zeta_A^0 = \biggl[ A + \left(A_g + A_l T_F n_l + A_h T_F n_b +
A_{bc}\left(\frac{m_c}{m_b}\right) T_F n_c \right)
\frac{\alpha_{s0}}{\pi} \Gamma(\varepsilon) m_{b0}^{-2\varepsilon}
+ \cdots \biggr]
\nonumber\\
&&\hphantom{\zeta_A^0={}}\times
C_F T_F n_b
\left(\frac{\alpha_{s0}}{\pi} \Gamma(\varepsilon) m_{b0}^{-2\varepsilon}\right)^2\,,
\label{j:A}
\end{eqnarray}
where $A_{bc}$ comes from Fig.~\ref{F:h}b, and
\begin{equation}
A_{bc}(1) = A_h\,,\quad
A_{bc}^{\text{hard}}(x\to0) = A_l\,,\quad
A_{bc}^{\text{hard}}(x\to\infty) \to A_s x^{-6\varepsilon}
\label{j:Abc}
\end{equation}
(in the hard region all loop momenta are of the order of the heaviest quark mass).
Combining the result for $A_{bc}$ with the previously known
single-scale results, we obtain the renormalized decoupling coefficient.

\section{Conclusion}
\label{S:Conc}

We have considered simultaneous decoupling of $b$ and $c$ quarks
with three-loop accuracy
for the fields and parameters of QCD Lagrangian~\cite{GHHS:11}
($\alpha_s$, light-quark masses $m_q$, gauge parameter $a$,
fields $A$, $c$, $q$),
as well as for bilinear quark currents~\cite{G:12}
(flavour non-singlet and singlet light-quark currents,
heavy-quark currents).
The coefficients of the leading power correction to the three-loop term
$(m_c/m_b)^2 (\alpha_s/\pi)^3$ happen to be of order 1,
while the coefficients of the four-loop corrections~\cite{SS:06,CKS:06}
$(\alpha_s/\pi)^4$ are numerically large.
Therefore, our power corrections are numerically negligible
(but this conclusion could not be obtained without calculating
these power corrections).

\textbf{Acknowledgements}. A.G.\ is grateful to DESY Zeuthen and TTP Karlsruhe
for financial support which allowed to attend the conference.
This work was supported by the BMBF through Grant No. 05H09VKE.

\end{document}